\documentclass[a4paper]{article}
\usepackage{graphicx}
\usepackage[T1]{fontenc} 
\usepackage{float}       
\usepackage{helvet}      
\usepackage{mathptmx}    
\usepackage{setspace}    
\AtBeginDocument{\doublespacing}
\newcommand{\degree}{\char'27\kern-.3em\hbox{C}}

\author{T. Mizoue$^{1}$, Y. Aoki$^{1}$, M. Tokita$^{2}$, H. Honjo$^{1}$, \\
H. J. Barraza$^{3}$, and H. Katsuragi$^{1}$\thanks{E-mail: \texttt{katsurag@asem.kyushu-u.ac.jp}} \\
\\
$^{1}$Department of Applied Science for Electronics and Materials, \\Kyushu University, Kasuga, Fukuoka 816-8580, Japan \\
$^{2}$Department of Physics, Kyushu University, \\Hakozaki, Fukuoka 812-8581, Japan \\
$^{3}$Unilever R{\&}D, Quarry Road East, Bebington CH63 9HW, \\United Kingdom \\
\\
}

\title{Control of polymer gel surface pattern formation and its three dimensional measurement method}

\begin{document}

\maketitle

\begin{abstract}
We report the controllability of a gelation surface pattern formation. Recently, we have found and studied a novel kind of pattern formation that occurs during a radical polymerization (gelation) process. The pattern formation is observed in an open top boundary of quasi two dimensional gelation. In previous studies, we have used two dimensional photo based image processing to analyze the patterns. However, the actual pattern is a three dimensional surface deformation. Thus we develop a three dimensional measurement system using a line laser displacement sensor and an automatic x-stage. Patterns measured by the system are analyzed and discussed by means of pattern controllability. In particular, we focus on the possibility of the pattern control using an external temperature field. As a result, we reveal that the global structure can be controlled, whereas the characteristic length scales (wavelength and amplitude) are not controllable.
\end{abstract}


\section{Introduction}

Pattern formation is one of the most fascinating and challenging problems in natural science. Complexity of nature has been making various kinds of beautiful patterns. We are surrounded by such beautiful patterns, e.g., honeycomb structure in a convection system \cite{Cross1993}, diffusion limited aggregation pattern \cite{Matsushita1984}, self-affine mountain profiles \cite{Matsushita1989,Katsuragi1999}, and so on \cite{Cross1993}. Pattern formation phenomena can be observed not only in physical systems, but also in chemical, and biological systems. For instance, our human body structure is a result of highly controlled pattern formation. Even in our everyday life, fluid system usually makes beautiful patterns such as a milk-crown structure \cite{Yarin2006}. In order to understand the complex nature of such pattern formations, more comprehensive studies have to be carried out. 
Especially, the pattern formation of soft matter is a growing research field since it relates to various scientific and industrial issues. Here, we focus on the pattern formation of polymer gel. Polymer gel is a typical example of soft materials which shows complex rheological property.

It is well known that the pattern formation is caused by a volume phase transition of polymer gel. Since the pattern formation was reported by Tanaka group \cite{Tanaka1987,Matsuo1992}, it has been studied both by experimental and theoretical ways \cite{Tokita1999,Tokita2000,Hwa1988,Onuki2002}. In this phenomenon, the origin of pattern formation is thought a mechanical instability owing to the abrupt volume change. 

In contrast, we have found a novel pattern formation appearing on a gelation surface \cite{Katsuragi2006}. The pattern itself looks similar to the previous (volume phase transition based) one. However, physical mechanisms behind them are completely different. The pattern formation we found occurs without volume phase transition. Furthermore, no other sudden change can be observed during pattern formation. 

Poly-acrylamide (AA) gel slabs have been mostly used to make the pattern. Poly-AA gel is very popular for biologists because they often use it for electrophoresis. Radical chain reaction (polymerization) governs the gelation of poly-AA.

We found that ambient oxygen is necessary to make the pattern \cite{Katsuragi2006,Mizoue2009}. The pattern formation occurs when the top of the sample is open to ambient air. Then oxygen is absorbed from a surface of slab, and diffuses in pre-gel solution during polymerization process. When we decrease the concentration of ambient oxygen, the pattern becomes thin \cite{Mizoue2009}. The oxygen is known as an inhibitor in the radical polymerization (gelation). The necessity of inhibitor in a diffusion system reminds us of the Turing instability. It is caused by diffusion and nonlinear competing reactions between an activator and an inhibitor. Such system is called Reaction Diffusion (RD) system. The importance of inhibitor (oxygen) suggests the system might be governed by the RD system. In addition, the typical length scale (wavelength of the pattern) were measured and analyzed based on the framework of RD system. Then, the length scale was consistent with the RD system \cite{Katsuragi2006}. 

In our previous study \cite{Mizoue2009}, we demonstrated the universality of this kind of pattern formation. We found that dimethylacrylamide (DMAA) and sodium acrylate (SA) polymers are also able to create surface pattern formation under similar experimental conditions. Thus we think this pattern formation is universal at least for quasi 2-dimensional (2D) free surface radical polymerization (gelation) process. 

We have measured the Effective Surface Roughness (ESR) using 2D photo data of the pattern \cite{Mizoue2009}. The ESR characterizes the degree of pattern formation. It indicates a large value for vivid pattern formation, and a small value for slight pattern formation. We varied three essential parameters (oxygen concentration, initiator concentration, and temperature), and measured the ESR of resultant patterns in order to examine the parameters dependence of the pattern formation. Finally, we obtained an empirical scaling which combines three parameters into a single scaling parameter \cite{Mizoue2009}. 

\begin{figure}
\begin{center}
\scalebox{0.7}[0.7]{\includegraphics{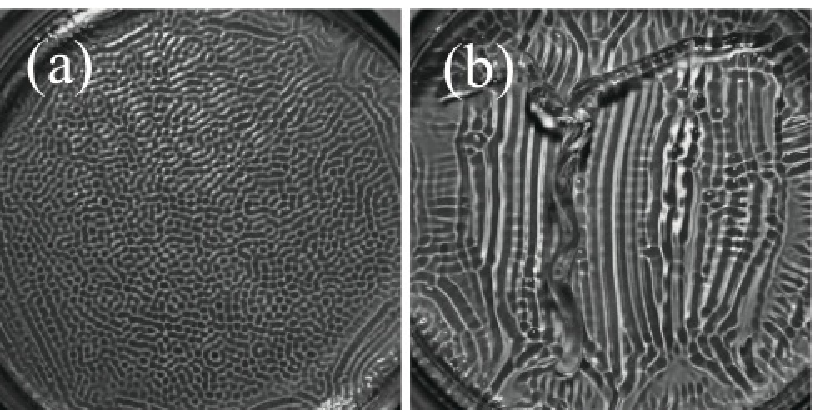}}
\caption{Typical patterns of poly-AA gel slabs made on $9$ cm Petri-dishes. (a) A random pattern appears in low temperature regime, and (b) a stripe pattern appears in high temperature regime \cite{Katsuragi2006}.}
\label{fig:rand_stripe}
\end{center}
\end{figure}

Although the actual pattern formation is 3-dimensional (3D) surface deformation, all analyses mentioned above are based on 2D photo data. Typical photos of the surface patterns are shown in Fig. \ref{fig:rand_stripe}. As shown in Fig. \ref{fig:rand_stripe}, 3D deformation can be visualized by 2D gray scale pattern by an inclined illumination. However, 3D direct measurement should be applied for the real quantitative analysis of 3D deformation. One main focus of this paper is to introduce a 3D measurement system to analyze the pattern formation directly. A line laser displacement sensor and an automatic x-stage are used for this purpose.

There is another research focus concerning the ordering of pattern formation. In Fig. \ref{fig:rand_stripe}, two kinds of patterns are shown, (a) random and (b) stripe. As reported in the previous paper \cite{Katsuragi2006}, random pattern appears in relatively low temperature regime. And stripe pattern appears in relatively high temperature regime. Usually, random structure should be observed in high temperature regime, since temperature is a source of thermal noise. In this sense, it is a counterintuitive result. The reason of this trend has been left unsolved. 


Very recently, we realized the substrate rack of the apparatus (constant temperature chamber), which is underneath the sample, has stripe shape. Besides, the pattern aligns to that direction. Therefore we guess that the stripe pattern might come from the substrate based external temperature field. In order to check this assumption, we use a heating wire array, and make a poly-AA gel slab on it. Direction, wavelength, and amplitude of the resultant patterns are measured by the 3D measurement system, and controllability of them are discussed in this paper. 

Finally, we investigate more general possibility of pattern control, using various shapes of external temperature field.

\section{Experimental}
\subsection{Materials}

The materials and methods to make gel slabs are basically same as previous studies \cite{Katsuragi2006,Mizoue2009}. AA monomer ($M_w$ = $71.08$) constitutes sub-chains, and methylenbisacrylamide (BIS, $M_w$ = $154.17$) constitutes crosslink. Ammonium persulfate (APS) is used as an initiator of radical reaction, and tetramethylethlyenediamine (TEMD) is used as an accelerator of the radical polymerization. Most of produced gels are composed by $1.2$ g AA, 6 mg BIS, $7$ - $10$ mg APS, $30$ $\mu$l TEMD, and $12$ ml deionized water.

The experimental procedure is very simple. Pre-gel solution is made in a beaker and degassed under ultrasonic vibration. Right after this procedure, the initiator is added and the solution is poured onto a Petri-dish ($9$ cm in diameter). Then surface deformation (pattern formation) occurs spontaneously within about 2 hours. This pattern remains permanently after the gelation. Typical thickness of gel slabs is approximately $2$ mm. The gel slabs are thin enough to be regarded as quasi-2D. 

\subsection{3D measurement system}
\begin{figure}
\begin{center}
\scalebox{1}[1]{\includegraphics{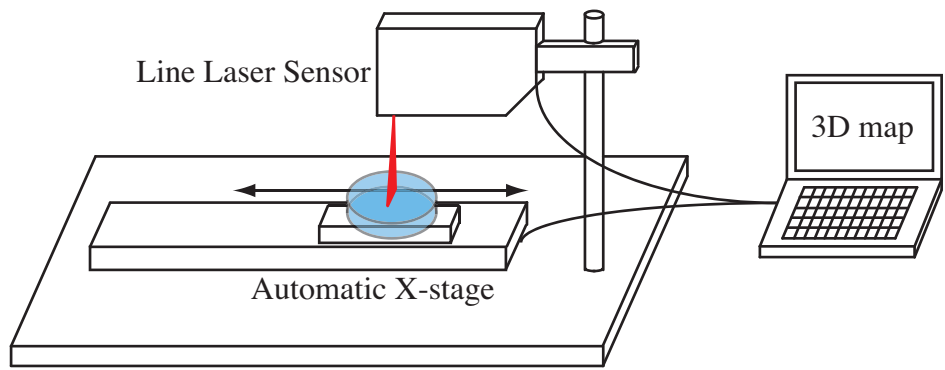}}
\caption{Schematic image of the 3D measurement system. Sample is held on an automatic x-stage, and a line laser sensor is mounted above it. Complete 3D surface height map can be obtained by the collection of intersectional profiles at different x positions.}
\label{fig:surface3d}
\end{center}
\end{figure}

We develop a 3D measurement system to directly measure the surface deformation of gel slabs.  Schematic image of the system is shown in Fig. \ref{fig:surface3d}. A line laser sensor (KEYENCE LJG-030) is mounted above a sample. This sensor is composed by a line laser unit and a 2D CMOS detector so that the sensor is able to acquire an intersectional profile of a 3D deformed surface. The sample is held on an automatic x-stage (COMS PM80B-100X). The axis of this stage is placed perpendicular to the line of laser. Sample is shifted by the x-stage after taking every intersection data. All instruments are controlled by a PC (with LabVIEW 8.6). Collected data are reorganized to the 3D surface profile. This system has $33$ $\mu$m horizontal resolution (both of x- and y-direction), and $10$ $\mu$m vertical resolution. Recently, the system is also applied to measure the drop-granular impact cratering \cite{Katsuragi2010}.


\section{Results}

\begin{figure}
\begin{center}
\scalebox{1}[1]{\includegraphics{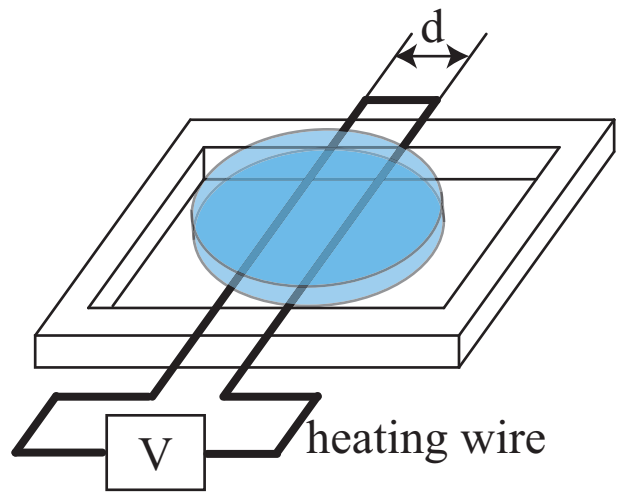}}
\caption{A heating wire array. The wire has $2$ mm diameter, but the contact is almost linear since the wire surface is rigid. In other words, the contact is 1-dimensional. The $d$ is interval distance between two successive wires. Temperature of the wire can be controlled by applying voltage to the wire.}
\label{fig:wires_array}
\end{center}
\end{figure}

First, we make a gel slab on a heating wire array as shown in Fig. \ref{fig:wires_array}. Temperature of the array is set as $30$ or $40\,$ {\degree} and environmental (room) temperature is $25\,$ {\degree}. When we make a gel slab on it, clear stripe structure can be observed. Direction of the stripe completely agrees with that of the wire array. This means that the direction of stripe pattern can be controlled by an external temperature field substrate. This is a qualitative check of the pattern controlability.


\subsection{Wavelength measurement}
\begin{figure}
\begin{center}
\scalebox{0.7}[0.7]{\includegraphics{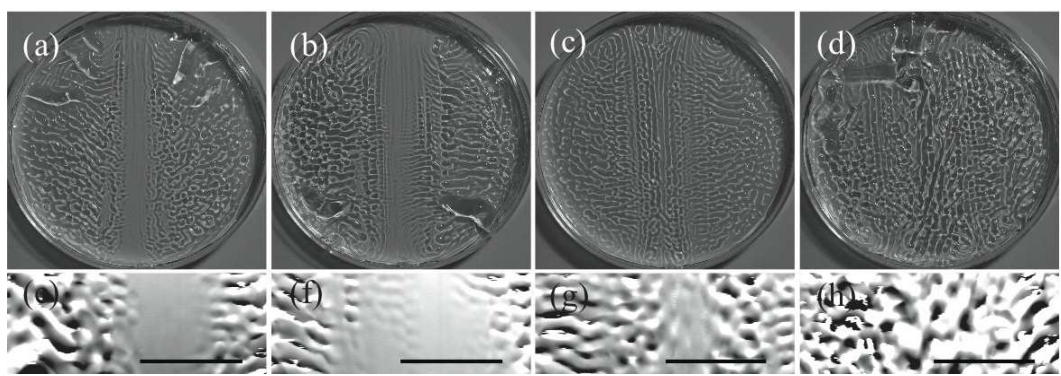}}
\caption{Typical patterns created on a heating wire array substrate. Photographs (upper) and 3D surface maps (lower) show following $d$ (distance between wires) cases respectively, (a,e) $d = 5$ mm, (b,f) $d = 10$ mm, (c,g) $d = 20$ mm, and (d,h) $d = 30$ mm. In all cases, wire temperature is set as 30 {\degree}, and initiator amount is 9 mg. Scale bars correspond to $10$ mm.}
\label{fig:hw_3d}
\end{center}
\end{figure}

Next, we are going to try to control the characteristic length scale (wavelength) of the stripe pattern. In this experiment, interval distance between two wires $d$ is a main control parameter. We varies $d$ as, $d$ = $5$, $10$, $20$, and $30$ mm. Temperature of the wire is fixed $30$ or $40\,$ {\degree}. Typical patterns (photos) and 3D measured results are shown in Fig. \ref{fig:hw_3d}.

3D measurement data corresponds to central $25 \times 10$ (mm$^2$) region of samples. As seen in Fig. \ref{fig:hw_3d}, the surface becomes rather flat with slight stripes when the interval distance is small. By contrast, independent stripe patterns appear if the interval distance is large enough. However, it is hard to confirm the correspondence between the interval distance and the pitch of stripe patterns. In other words, we are not able to see clear evidence of wavelength controllability in these data. 

\begin{figure}
\begin{center}
\scalebox{1}[1]{\includegraphics{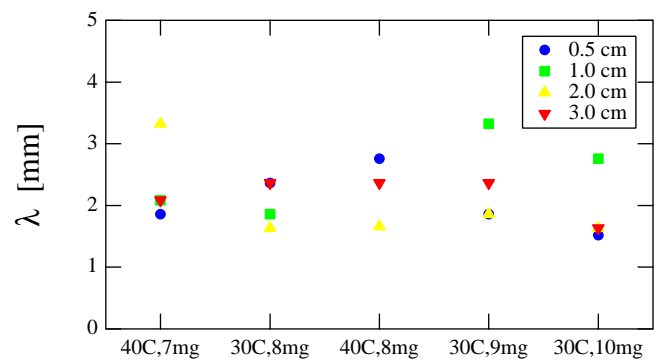}}
\caption{Characteristic length scale (wavelength of the stripe pattern) of the poly-AA gel slabs with different interval distance of heating wires. FFT analysis is applied to 3D data to compute the wavelength. Symbols indicate heating wires interval $d$. Other experimental conditions (wire temperature and initiator amount) are written in horizontal axis label.}
\label{fig:lambda}
\end{center}
\end{figure}

Using 3D height map data, we now analyze the characteristic length scale of the patterns more accurately. We apply the Fast Fourier Transform (FFT) analysis to the 3D data (Fig. \ref{fig:hw_3d}(e-h)). The peak wave number obtained by FFT analysis is translated to the wavelength $\lambda$ (characteristic length scale). Measured $\lambda$ is shown in Fig. \ref{fig:lambda}. Horizontal axis in Fig. \ref{fig:lambda} indicates experimental conditions (wire temperature and initiator amount). Symbols represent the interval distance of heating wires as denoted in the legend. In Fig. \ref{fig:lambda}, we cannot confirm clear trend of the wavelength data. It seems to fluctuate around an average value of $2$ mm. This value certainly agrees with the previous result \cite{Katsuragi2006}.


The result suggests that it is hard to control the wavelength of the patterns. The stripe patterned external temperature field makes a stripe patterned gel slab. However, it is impossible to control the wavelength (pitch) of the pattern. In the previous paper \cite{Katsuragi2006}, we considered that the origin of the pattern formation is a sort of RD based instability. If it is true, the characteristic length scale (wavelength) should be determined by the diffusion length scale of the inhibitor \cite{Walgraef1997}. In RD system, the wavelength is intrinsically determined rather by internal conditions than by external conditions. The result obtained here is consistent with that consideration. Only the direction can be controlled by the external temperature field.

\subsection{Amplitude measurement}
\begin{figure}
\begin{center}
\scalebox{1}[1]{\includegraphics{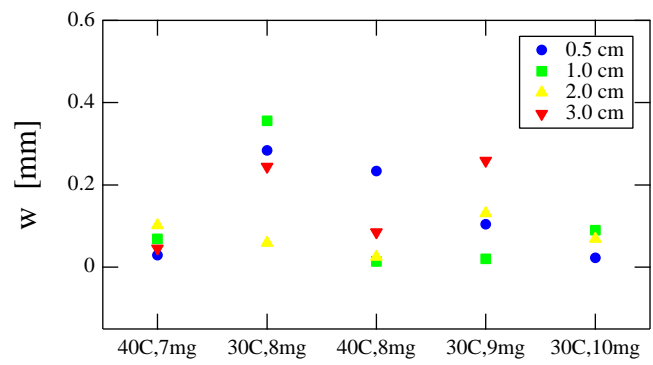}}
\caption{Surface roughness (standard deviation of 3D height map data) obtained from the same data as those in Fig. \ref{fig:lambda}. Symbol code and horizontal axis label are also same as those in Fig. \ref{fig:lambda}.}
\label{fig:w}
\end{center}
\end{figure}

Next, we measure the amplitude of surface deformation. The surface roughness $w$ (standard deviation of height map data) is computed on 3D data that are same as those used in Fig. \ref{fig:lambda} computation. The result is shown in Fig. \ref{fig:w}. Although the data scatter a lot, there is not any clear trend, again. The characteristic amplitude order (surface roughness) is about $10^{-1}$ mm. 

According to the data in Figs. \ref{fig:lambda} and \ref{fig:w}, controllability of the pattern formation is very limited. We are able to control global structure (direction) of the pattern, whereas both of the wavelength and amplitude controls are impossible. The length scales seem to be determined by other physical mechanisms like diffusion of inhibitor, and/or polymerization dynamics, etc. 


\subsection{Global structure control}

Here, we try a little bit more challenge of the pattern control. Since global structure of the pattern formation is controllable, we make gel slabs on several shapes of external temperature fields. The fields are generated by heated metal molds that have square, circle, and heart shapes. The metal molds are heated by a heating wire binding the molds.

\begin{figure}
\begin{center}
\scalebox{0.8}[0.8]{\includegraphics{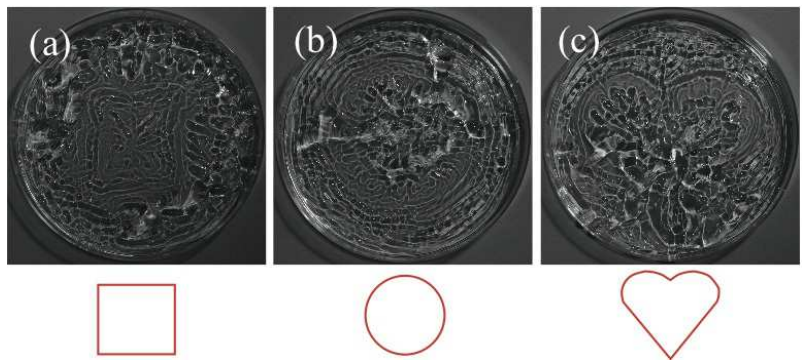}}
\caption{Pattern control by some shapes of external temperature field: (a) square, (b) circle, and (c) heart.}
\label{fig:molds}
\end{center}
\end{figure}

In Fig. \ref{fig:molds}, photos of the obtained gel slabs and corresponding temperature field shapes are displayed. As expected, we can control the global structure of resultant patterns. They are clearly affected by external temperature fields. Furthermore, the wavelength seems to be independent of the temperature field. These characteristics are consistent with the result obtained by the wire array experiment. 


\section{Discussion}

The obtained results so far conclude the difficulty of the length scale control. This pattern formation phenomenon itself is very robust. The fine parameter tuning is not necessary to make the pattern. In this sense, the pattern formation is a kind of self-organized phenomenon. It is hard to control the length scale due to the robustness of self-organization. On the other hand, we can easily control the global structure of the pattern formation. Surface pattern deformation simply obeys external temperature field shape. 

In Ref.\cite{Katsuragi2006}, we considered the transition from random to stripe patterns at a higher temperature region. In the previous experiments, temperature is controlled by a constant temperature chamber with a stripe shaped substrate rack. Thus the environmetal temperature was controlled. In low temperature reigme, heat condution and gelation process is rather uniform. Contrastively, heat conduction from the substrate rack cannot be negligible in high temperauture regime. That is a reason of transition like behavior at a certain temperature. In the current experiment, we use a heating wire array to actively control the temperature field, and the environmental temperature is room temperature ($\sim 25\,$ {\degree}). As a result, the created pattern obeys the temperature field. Therefore, we think that the temperature gradient plays an essential role in this phenomenon. However, the precise measurement of the temperature field is not easy. The current temperature control system is too rough to measure and discuss such a delicate temperature field. The precise temperature field measurement is a future problem.

The typical length scale of the pattern formation is in an order of $10^{-1}$-$10^0$ mm. Such macroscopic self-organized pattern is not so common in polymer science. In this report, we do not discuss the microscopic dynamics of polymerization, or microscopic polymer network structure. The relation between them (macroscopic pattern and microscopic dynamics/structure) is an open problem.
Numerical model may be helpful to understand the physics of this pattern formation. However, it will not be so easy. Obviously, diffusion coefficients of monomer, growing polymer blobs, and oxygen, decrease drastically during polymerization (gelation). And finally they will be almost arrested by the elasticity of gel. It is difficult to model such time dependent diffusion problem.  In addition, large amount of memory must be needed to reproduce macroscopic pattern formation from microscopic polymerization dynamics model. Macroscopic phenomenological model might be suitable for this type of problem. 

Moreover, macroscopic pattern formation of soft matter might relate to the patterning problem of the bio-materials, such as brains, reptiles skin, and so on. Wrinkling on old human's skin is also a kind of pattern formation. The pattern formation reported here is more or less similar to such bio-related pattern formation. But it just looks similar. It never means underlying physics conformity. We cannot conclude anything about this similarity at this moment.

\section{Conclusion}

We have examined the controllability of the spontaneous pattern formation occurring on a gelation surface. A heating wire array was utilized to make external temperature field. Pre-gel solution was put on it during gelation. Then the ordered structure that aligns to the array direction appeared. This implies that we can control the global structure of the pattern formation. We next varied interval of heating wires and put pre-gel solution on it to investigate the controllability of the length scale (wavelength and amplitude) of the pattern formation. These surface patterns were measured by a 3D surface deformation measurement system. That system is composed by a line laser sensor and an automatic x-stage and has $10^1$ $\mu$m resolution. 

From the data analyses, we found that both of wavelength and amplitude cannot be controlled by external temperature field. These length scales seem to be determined by internal parameter such as diffusion length scale.

\bibliographystyle{plain}

\clearpage

\clearpage

\clearpage

\clearpage

\clearpage

\clearpage

\clearpage

\end{document}